\documentclass[aps,prl,reprint,floatfix,showpacs]{revtex4-1}
\usepackage{amssymb,amsmath}
\usepackage{graphicx}

\usepackage[scanall]{psfrag}

\usepackage{color}

\newcommand{\abs}[1]{\left| #1 \right|}
\newcommand{\Eqref}[1]{\eqref{#1}}

\graphicspath{{./Figs/fraggedFigs/}}

\begin{document}

\title{Using the uncertainty principle to design simple interactions for targeted self-assembly}

\author{E. Edlund}
\author{O. Lindgren}
\author{M. \surname{Nilsson Jacobi}}%
 \email{mjacobi@chalmers.se} 
\affiliation{Complex Systems Group, Department of  Energy and Environment, Chalmers University of Technology, SE-41296 G\"oteborg, Sweden}
\date{\today}%

\begin{abstract}
We present a method that systematically simplifies isotropic interactions designed for targeted self-assembly.
The uncertainty principle is used to show that an optimal simplification is achieved by a combination of heat kernel smoothing and Gaussian screening.
We use this method  to design isotropic interactions for self-assembly of complex lattices and of materials with functional properties.
The interactions we derive are significantly simpler than those previously published, and it is realistic to discuss explicit experimental implementation of the designed self-assembling components.
\end{abstract}


\maketitle

%%%%%%%%%%%%%%%%%%%%%%%%%%%%%%%%%%%%%%%%%%
% Introduction
%
Fabrication of functional materials through directed self-assembly is an attractive possibility for many reasons, such as scalability, robustness, flexibility, and cost efficiency. 
Two main challenges must be addressed before this idea becomes useful in practice. 
First, we must be able to determine the necessary characteristics of the self-assembling building blocks.  
There are several approaches to this: extensive computational explorations~\cite{mossa_ground-state_2004,zhang_self-assembly_2004,Glotzer:2004ce,engel_self-assembly_2007}, 
problem specific heuristics~\cite{spillmann_hierarchical_2003,chen_directed_2011,sinclair_generation_2011,macfarlane_nanoparticle_2011}, and, as in this paper, theoretical methods~\cite{cohn_algorithmic_2009,hormoz_design_2011,torquato_inverse_2009,edlund_designing_2011}.
Second, we must be able to manufacture the components according to the derived design. 
This second point is problematic.
So far the theoretically designed components rely on interactions that are too complicated to be implemented experimentally.

The gap between theory and experiment is continuously reduced by the rapid advances on the experimental side
where it is now possible to fabricate nano- and mesoscale particles with complex and tunable interactions~\cite{min_role_2008,bishop_nanoscale_2009,castro_primer_2011,bianchi_patchy_2011}. 
While these results are impressive in themselves, the gap between theory and experiments remains wide and is unlikely to be closed from the experimental side alone.
In this paper we report progress from the theory side, bringing analytically designed systems closer to experimental realization.

In this Letter we introduce a general principle for how to systematically simplify isotropic interaction potentials that lead to self-assembly of metamaterials with desired properties. 
The  method is based on the uncertainty principle, which limits how short-ranged and smooth a potential can be without losing the features in reciprocal space that guarantee the self-assembly of the target structure. 
A consequence of the uncertainty principle is that heat kernel smoothing of the reciprocal potential is optimal in the sense that the interaction range decreases  at a minimal loss of detail in the reciprocal (design) space.

We first illustrate the simplification method by designing a self-organized disordered material with suppressed diffraction around a chosen frequency, i.e., with near invisibility in the chosen frequency range. 
The second example, our main result, are interaction potentials that self-assemble into 2D Kagome and 3D diamond lattices and are much simpler than previously published interactions.
The simplifications are significant enough to allow for speculation on possible experimental implementations.

%%%%%%%%%%%%%%%%%%%%%%%%%%%%%%%%%%%%%%%%%
%\section{Targeted self-assembly} 

As a model we consider a system of spherically symmetric (colloidal) particles that self-assemble by minimizing the potential energy.  
A configuration is described by its density, expressed as a sum of Dirac delta functions centered at the particle positions ${\bf r}_i$, $\rho ( {\bf r} ) = \sum _i \delta ( {\bf r} - {\bf r}_i )$. 
The energy is defined by a pairwise isotropic potential $V (r)$ as
\begin{equation}
E = \int d{\bf r} d {\bf r}' V(\abs{ {\bf r} - {\bf r}'}) \rho ( {\bf r} ) \rho ( {\bf r} ' ) = \int d k  \widehat{V} ( k ) \abs{ \hat{\rho} (k) }^2 
\label{energy}
\end{equation}
where $\abs{ \hat{\rho} (k) }^2  dk =  \int _{| {\bf k } | = k } d {\bf k } \abs{ \hat{\rho} ({\bf k })}^2$ and $\widehat{V} (k)$ is the Fourier transform of the (spherically symmetric) potential~\cite{Folland97}. In~\cite{edlund_designing_2011} it was demonstrated that \Eqref{energy} can be used to design potentials for directed self-assembly of crystal structures by penalizing competing configurations. However, direct implementations of such schemes typically result in complicated long ranged interactions.

%%%%%%%%%%%%%%%%%%%%%%%%%%%%%%%%%%%%%%%%%%%%
%\section{Simplifying designed interactions} 

{\bf Simplifying designed interactions.} This paper focuses on simplification of designed interaction potentials. Arguing that a potential is simple requires a definition of complexity.
In practice such a definition depends on what types of interactions can be realized experimentally, and is therefore problem specific and not immediately useful in a general setting.
Still, there are generic types of complications that cause problems for experimental implementations, such as potentials with many extremal points or with abrupt changes in their derivatives.
One suggested approach for moving towards realizability is to consider only convex interaction potentials~\cite{cohn_algorithmic_2009,marcotte_optimized_2011}. 
Convexity avoids the problem of multiple maxima and minima in the potential but instead moves the complexity into its derivatives. which is not necessarily easier to implement in experiments.

Rather than trying to find general requirements for realizability we focus on simplification of a given potential, which we argue is a more tractable problem since it is not based on an absolute measure of interaction complexity \footnote{Defining good measures of complexity is surprisingly hard and the subject has a long and contentious history~\cite{badii_complexity:_1999,lloyd_complexity_1988,Crutchfield:1989tn}. 
None of the existing measures pertains to complexity in interaction potentials and we doubt one will be found.
Even if we could define a quantitative measure with desirable theoretical properties, it is unlikely it will agree with realizability in the wide range of experimental systems available today. 
}.

Recall that  \Eqref{energy} can be used to design the potential in reciprocal space. Combining the reciprocal design principle with  the potential's physical realization in real space shows that the uncertainty principle is applicable. 
This places limits on the simultaneous localization of $V(r)$ and $\widehat{V} (k)$.   
To see in detail how the uncertainty relation works in our context, we start by defining a {\it screening} of a potential as a transformation $V(r) \rightarrow \xi (r) V(r)$, where $\xi$ is a positive function  monotonically decaying away from the origin. For technical reasons we also assume that $\xi$ can be normalized, $\int dr r \xi (r) ^2 = 1$. The effect in reciprocal space of the multiplicative screening is defined by a convolution: $\mathcal{F} \left[ \xi (r) V(r) \right] = \int d {\bf k}' \hat{\xi} ( | {\bf k} - {\bf k}' | ) \widehat{V} ( | {\bf k} | )$.
The standard uncertainty relation gives a central result that limits the dispersion around zero of $\xi$ and its Fourier transform $\hat{\xi}$ in $d$ dimensions~\cite{Folland97},
			
\begin{equation}
\operatorname{var} [ \xi ^2 ] \operatorname{var} [¬¨‚Ä†\hat{\xi}^2 ] \ge \frac{d^2}{16 \pi ^2} .
\end{equation}
The lower bound is achieved exactly when $\xi(r)$  is a Gaussian function, which defines a heat kernel smoothing in reciprocal space.
The dispersion $\operatorname{var} [ \xi^2]$  measures the degree of screening achieved by the transformation $V(r) \rightarrow \xi(r) V(r)$, whereas $\operatorname{var} [ \hat{\xi}^2]$ defines the loss of distinction by the corresponding kernel convolution. 
The interpretation of the uncertainty relation is therefore that a heat kernel smoothing in the reciprocal space achieves minimal loss of distinction for a given degree of screening of the potential in real space, and vice versa.

Based on this argument, we propose the following operation to simplify a designed potential:
\begin{eqnarray}
\widehat{V} ( k) & \xrightarrow{\mbox{smooth}} & \widehat{V} _{\tau} (k) = \int d {\bf k}' e ^{- | {\bf k} - {\bf k}' |^2 / \tau } \widehat{V} ( | {\bf k} ' | ) .
\label{heat_kernel}
\end{eqnarray}
This results in a screening of the potential in real space
\begin{eqnarray}
V(r)  & \xrightarrow{\mbox{screen}} &  V_{\tau} (r) =  \mathcal{F}^{-1} \left[ \widehat{V}_{\tau} \right] =  e ^{- \tau r^2 /4 } V(r)  ,
\label{screening}
\end{eqnarray}
where we ignore the irrelevant normalizing scale factors.
The role of $V(r)$ and $\widehat{V}(k)$ can also be reversed, i.e., a screening in reciprocal space results in a smoothing in real space. 
Both these operations reduce the complexity of the interaction potential and they will be used in combination later in the demonstration with self-assembling lattices.
The conclusion is that Eqs.~\eqref{heat_kernel}, \eqref{screening}, and their reciprocal duals define a general method for simplification of a designed isotropic interaction. 

We use Gaussian screening and smoothing because it is optimal in the sense defined by the lower bound in the uncertainty principle. In practice other types of screening may be easier to implement, for example an exponential damping $\xi (r) = e ^{-\tau r}$ like that of the Coloumb interaction between charged colloids due to free counterions. It is straight forward to generalize Eqs.~\eqref{heat_kernel} and \eqref{screening} by considering other smoothing kernels and their radial Fourier transforms as screening functions.

\begin{figure*}[htb]
\centering
%\psfrag{A}[c][c][1.2][0]{\textcolor{white} A}
%\psfrag{B}[c][c][1.2][0]{\textcolor{white} B}
%\psfrag{C}[c][c][1.2][0]{\textcolor{white} C}
%\psfrag{D}[c][c][1.2][0]{\textcolor{white} D}
%\psfrag{E}[c][c][1.2][0]{\textcolor{white} E}
%\psfrag{F}[c][c][1.2][0]{\textcolor{white} F}
%
%\psfrag{Vk}[c][c][\psSymbolSize][0]{$\widehat{V}(k)$}
%\psfrag{k}[c][c][\psSymbolSize][0]{$k$}
%\psfrag{kappa}[c][c][\psSymbolSize][0]{$\kappa$}
%\psfrag{Vr}[c][c][\psSymbolSize][0]{$V(r)$}
%\psfrag{r}[c][c][\psSymbolSize][0]{$r$}
%\psfrag{1}[c][c][\psSymbolSize][0]{$1$}
\includegraphics[width= 1 \textwidth]{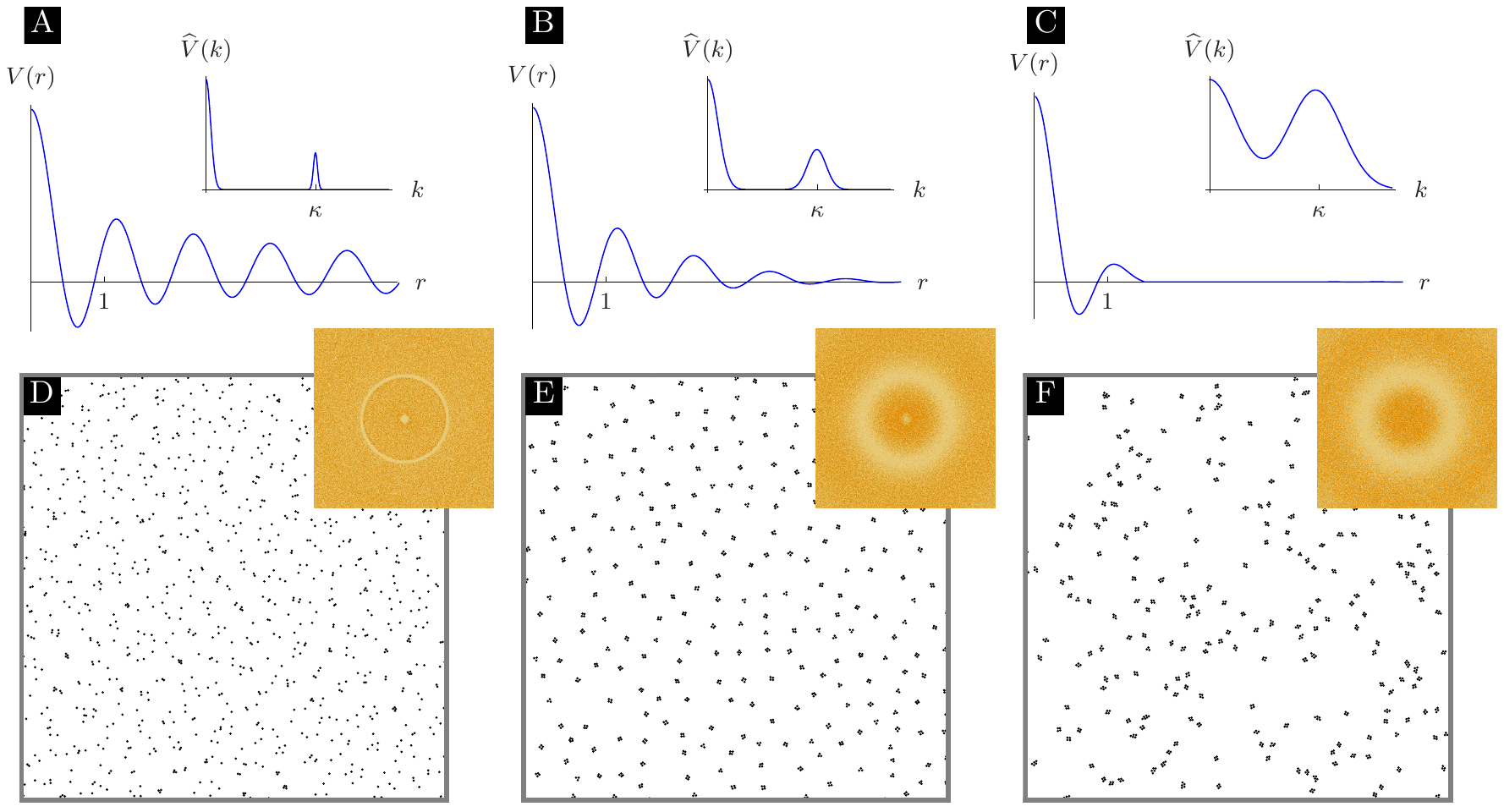}
\caption{\label{smoothedInteractions}
Heat kernel smoothing simplifying interactions for suppressed diffraction around a wavelength. 
A potential is constructed using~\Eqref{cylindricGauss} to penalize a certain wavelength $\kappa$ and the inverse Fourier transform gives the corresponding interactions, which have long-ranged oscillations (a). 
By applying the heat kernel smoothing in~\Eqref{heat_kernel}, we can successively simplify the interactions (b-c) until only a single minimum and maximum remains. 
%The peak at $\widehat{V} (0)$ is used to prevent aggregation.
(d-f) Low temperature configurations from Monte Carlo simulations verifying that the interactions cause self-organization into states with diffraction patterns with pronounced frequency gaps (insets show their diffraction patterns).
}
\end{figure*}

%%%%%%%%%%%%%%%%%%%%%%%%%%%%%%%%%%%%%%%%%
%\section{Example 1: Metamaterial with suppressed diffraction} 

{\bf Metamaterial with suppressed diffraction.} The most natural functional property for a self-assembling material is designed diffraction properties, since it is determined immediately by the assembled structure. The Fourier transform of a particle configuration (the structure factor), tells to what extent diffraction of incident light occurs in different directions. In this example we demonstrate metamaterials with suppressed structure factors within a desired frequency range, self-assembled from interactions of varying complexity. Since no scattering occurs for certain wavelengths in this material, it would essentially be invisible at those wavelengths~\footnote{Materials exhibiting this property have been called ''stealth-materials''~\cite{Torquato_stealth_2008}}.

From the expression of the energy in \Eqref{energy} it follows directly how to construct a potential that energetically disfavours configurations without suppressed structure factor: by letting $\widehat{V} ( k )$ be large for wave numbers $k$ in the bandwidth we want to penalize. 
Such potentials have previously been used in optimization algorithms to construct materials with similar properties \cite{florescu_designer_2009}. 
The functional form of the resulting potentials are linear combinations of a few Bessel functions that oscillate and have slow asymptotic decays, $V(r) \sim \cos (r) / \sqrt{r}$. 
They are therefore too complicated and long ranged to be useful in a self-assembly context, see Fig.~\ref{smoothedInteractions}a. 
This provides a natural setting for our method. We can apply the heat kernel smoothing in~\Eqref{heat_kernel} to $\widehat{V} (k)$ to construct interactions with significantly simpler functional forms. 
Here the trade-off is particularly transparent as the reduced localization in $\widehat{V} (k)$ makes the suppressed region of the structure factor less distinct.

To be more specific, consider a particle system in two dimensions for which we want to design a potential that results in a structure factor suppressed suppressed a specific wave length $\kappa$. Starting with $\widehat{V}_{0 , \kappa} (k)$ consisting of a Dirac delta function located at $\kappa$ and applying the heat kernel smoothing with parameter $\tau$ results a potential with reciprocal form  

\begin{equation}
\widehat{V}_{\tau , \kappa} (k) =  e ^{-(k^2 + \kappa ^2)/\tau} I_0 \left( 2 k \kappa / \tau \right) ,
\label{cylindricGauss}
\end{equation}
where $I_0$ is a modified Bessel function and $\kappa$ is the approximate location of the maximum (assuming $\kappa / \sqrt{\tau}$ is large), see Fig.~\ref{smoothedInteractions}a (inset).
From the uncertainty principle it follows that this construction leads to minimal range of the potential in real space for a given degree of localization around the maximum in reciprocal space. To prevent the particles from collapsing into a single aggregate we follow the idea in~\cite{florescu_designer_2009} and add a low frequency penalty to the energy spectrum \footnote{The effect of the penalty at low wave numbers was referred to as hyper-uniformity in~\cite{florescu_designer_2009}. }.

In Fig.~\ref{smoothedInteractions} different levels of smoothing are demonstrated together with the resulting configurations and their corresponding structure factors. 
Note that even with very strong screening, where the potential shows only a single over-damped oscillation, the suppression remains significant, as seen in the structure factor.

%%%%%%%%%%%%%%%%%%%%%%%%%%%%%%%%%%%%%%%%%
%\section{Example 2: Complex lattices} 

\begin{figure*}[t]
%\psfrag{A}[c][c][1.2][0]{\textcolor{white}A}
%\psfrag{B}[c][c][1.2][0]{\textcolor{white}B}
%
%\psfrag{U}[l][r][\psSymbolSize][0]{$\widehat{V}(k)$}
%\psfrag{k}[c][c][\psSymbolSize][0]{$k$}
%\psfrag{V}[b][c][\psSymbolSize][0]{$V(r)$}
%\psfrag{r}[c][c][\psSymbolSize][0]{$r$}
\includegraphics[width= 0.9 \textwidth]{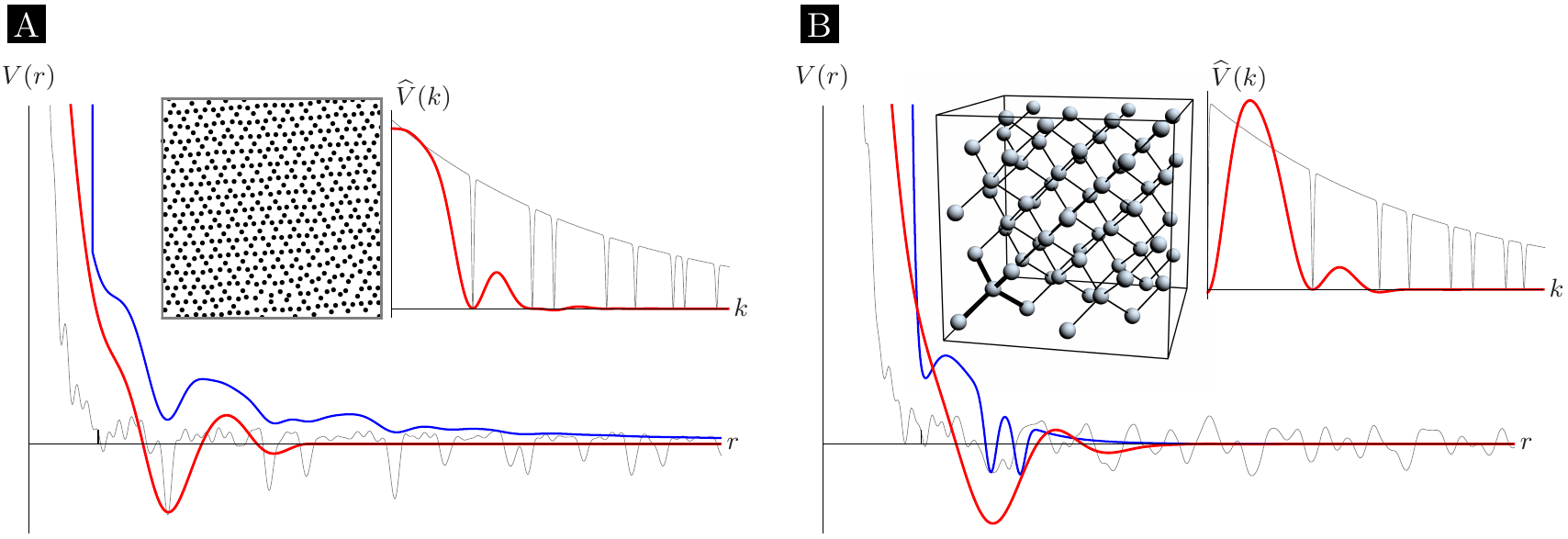}
\caption{\label{kagome} 
The main results: simplified interactions designed for lattice self-assembly.
The curves show the interactions from~\Eqref{lattice_construct} for targeted self-assembly in reciprocal space (inset) and in real space of a Kagome lattice (a) and a diamond lattice (b).
The gray curves are the basic construction with almost no screening or smoothing while  
the red curves show the optimized potentials after applying the method defined in~\Eqref{heat_kernel}.
As a reference we also show previously published potentials (blue curves) causing self-assembly of those lattices~\cite{edlund_designing_2011,Rechtsman07}.
}
\end{figure*}
{\bf Complex lattices.} One of the draws of directed self-assembly is the ability to design and implement microstructure displaying functional behavior, with self-assembly of for example chiral catalysts~\cite{ortega_lorenzo_extended_2000} and optical crystals~\cite{Jenekhe:1999ia} achieved to date. 
Thus much of the theoretical work in this area has concentrated on solving the inverse statistical-mechanical problem for low-coordinated crystals~\cite{torquato_inverse_2009}, i.e., designing interactions that lead to self-assembly of a target lattice, which is also our focus here.

Suppose we want to construct an isotropic potential that leads to self-assembly of a target lattice. 
The lattice can be described in reciprocal space by the structure factor  $| \hat{\rho} ( {\bf k}  ) |^2$, which is positive and has support only at discrete points ${\bf k}_i $. 
A basic construction of a potential that has the target lattice as a lowest energy state, with energy zero, is to let $\widehat{V} (k)$ be a positive function with zeros at all points $| {\bf k}_i |$. 
This scheme works in principle but implemented naively it leads to very complicated and long range interactions~\cite{edlund_designing_2011}, illustrated by the gray curves in Fig.~\ref{kagome}.
Starting from the basic design however, it is possible to use the heat kernel smoothing and screening to produce a much simplified potential that still causes assembly of the target lattice. 
Explicitly we use a construction on the form
\begin{equation}
	V_{\tau \sigma} (r) = e^{- \tau r^2} \mathcal{F} ^{-1} \left[ e ^{- \sigma k^2} \left( 1 -\sum _i \alpha _i \widehat{V}_{\epsilon ,  |{\bf k}_i|} (k)  \right)  \right] ,
	\label{lattice_construct}
\end{equation}
where $\widehat{V}_{\epsilon ,  |{\bf k}_i|}$ is defined in~\Eqref{cylindricGauss}%
~\footnote{\Eqref{cylindricGauss} is defined in two dimensions, the generalization to 3D is $\hat{V}_{\epsilon,|{\bf k}_i|}(k) = \frac{2 |{\bf k}_i|}{\sqrt{\epsilon \pi} k}e^{-(|{\bf k}_i|^2+k^2)/\epsilon}\sinh{(2|{\bf k}_i| k/\epsilon)}$.%
}  %
and the parameters $\tau$ and $\sigma$ controls the degree of screening in real and reciprocal space respectively, and $\epsilon$ is small.
 The parameters $\alpha _i$ and $|{\bf k}_i|$ are set so that $\widehat{V}_{\tau \sigma} (k)$ fulfills the basic design principle in~\cite{edlund_designing_2011}, i.e., $\widehat{V} _{\tau \sigma} ( | {\bf k} | ) \geq 0$ and $\widehat{V} _{\tau \sigma} ( | {\bf k}_i | ) =0$, with a small perturbation that breaks the energy degeneracy to favor the target lattice. The parameters $\tau$ and $\sigma$ should be chosen as large as possible without losing the target lattice as the lowest energy state.

In Fig.~\ref{kagome} we demonstrate the method on a Kagome lattice and a diamond lattice. 
The main results are two maximally simplified potentials that are shown to self-assemble into the target lattices in Monte Carlo simulations.
We compare the results to our previously published potential for the Kagome lattice, which was the simplest interactions we could achieve  with ad hoc tuning in~\cite{edlund_designing_2011}, and to interactions that gives a diamond lattice, derived by Rechtsmann et al.~\cite{Rechtsman07} using an iterative design scheme (blue curves in Fig.~\ref{kagome}). 
This indicates how effective the systematic approach we suggest is. 

%%%%%%%%%%%%%%%%%%%%%%%%%%%%%%%%%%%%%%%%%
%\section{Summary and discussion} 
{\bf Summary and discussion.} 
We have demonstrated how the uncertainty principle  can be used to design relatively simple isotropic potentials for targeted self-assembly. 
Our result also indicate fundamental reasons for why an isotropic potential must be relatively complicated to achieve self-assembly of a complex lattice. 

From the experimental side there is a continuous progression of techniques for producing nano- and mesoscale components with custom designed interactions and shapes. 
Colloids in particular have been identified as suited for tailoring interactions of the type we discuss in this article. 
Effective pair interactions between colloidal particles arising from e.g.\ charge, surface features, electrolytes or polymers in the solution, or dipole moments can be combined and tuned~\cite{likos_effective_2001,Yethiraj:2007if}. 
Complex potentials with many minima will likely be out of reach for the foreseeable future but the simplified potentials we present here are much more experimentally accessible.
For example, the diamond potential of Fig.~\ref{kagome}(b) could be implemented if a Coloumb repulsion could be combined with depletion attractions~\cite{roth_depletion_2000,*dijkstra_direct_1999}.
We believe that an implementation of this kind would be an important step towards practical realizations of designed functional materials.

OL and MNJ acknowledge support from the SuMo Biomaterials center of excellence. We also thank Kristian Lindgren for helpful discussions.  

%\bibliographystyle{apsrev4-1}
%\bibliography{opticalBandgap}


\end{document}